# Delay-Distance Correlation Study for IP Geolocation

□ **DING Shichang**[1], **LUO Xiangyang**[1], **YE Dengpan**[2], **LIU Fenlin**[1]†

1. The State Key Laboratory of Mathematical Engineering and Advanced Computing, Zhengzhou 450001, Henan, China;
2. School of Computer, Wuhan University, Wuhan 430072, Hubei, China

**Abstract:** Although many classical IP geolocation algorithms are suitable to rich-connected networks, their performances are seriously affected in poor-connected networks with weak delay-distance correlation. This paper tries to improve the performances of classical IP geolocation algorithms by finding rich-connected sub-networks inside poor-connected networks. First, a new delay-distance correlation model (RTD-Corr model) is proposed. It builds the relationship between delay-distance correlation and actual network factors such as the tortuosity of the network path, the ratio of propagation delay and the direct geographical distance. Second, based on the RTD-Corr model and actual network characteristics, this paper discusses about how to find rich-connected networks inside China Internet which is a typical actual poor-connected network. Then we find rich-connected sub-networks of China Internet through a large-scale network measurement which covers three major ISPs and thirty provinces. At last, based on the founded rich-connected sub-networks, we modify two classical IP geolocation algorithms and the experiments in China Internet show that their accuracy is significantly increased.

**CLC number:** TP 393

**Received date:** 2016-09-08
**Foundation item:** Supported by the National Natural Science Foundation of China (61379151, 61274189, 61302159 and 61401512), the Excellent Youth Foundation of Henan Province of China (144100510001) and Foundation of Science and Technology on Information Assurance Laboratory ( KJ-14-108)
**Biography:** DING Shichang, male, Ph.D. candidate, research direction: IP Geolocation. E-mail: dscgongzuo@yeah.net

† To whom correspondence should be addressed. E-mail: luoxy_ieu@sina.com

## 0 Introduction

IP geolocation can find the geographical location of a host based on its IP address. It is very valuable to many location-based applications (especially many location-based network security services): law enforcement organizations rely on IP geolocation to identify where cyber-attacks come from; online banks can design credit card fraud identification method based on users' locations; IP geolocation can also be used in designing Network Resource Access Control Protocol, etc. An accurate IP geolocation service is one of the most important foundations of location-based security services [1-4].

There are three kinds of IP geolocation techniques: IP geolocation databases, data-mining-based IP geolocation and delay-based IP geo-location. IP geolocation databases can hardly satisfy the demand of some applications because there are many conflicts between different databases [5,6]. Data-mining-based IP geolocation techniques may have to map an IP segment to one location when they cannot directly find the location of some IP addresses, and this could cause significant localization errors [7]. Delay-based IP geolocation techniques estimate the location of the target IP on the network delays from known landmarks to the target IP [4,8-11]. So theoretically they can directly geolocate any IP address. Though there are several street-level geolocation techniques [3,4], in this paper we mainly discuss about city-level delay-based IP geolocation for street-level geolocation is actually based on an accurate city-level IP geolocation service.

There are many classical delay-based IP geolocation algorithms which are suitable to rich-connected networks such as U.S. and Western Europe (the rich-connected network means delay-distance correlation of this network is strong while the poor-connected network means weak). The assumption these algorithms rely on is that the correlation between network delay and geographical distance is strong [8-11]. However, the delay-distance correlation of some networks like China Internet is too weak for these algorithms, which seriously affects their accuracy [11]. This paper observes that though delay-distance correlation of a poor-connected network seems weak as a whole, delay-distance correlation of some sub-networks inside the poor-connected network may be much stronger. It indicates that if we make sure probing hosts and landmarks are deployed in the same rich-connected sub-networks of target hosts, the performances of classical IP geolocation algorithms can be improved.

The problem is how to find the rich-connected sub-networks. There has been studies about delay-distance correlation. Wang Yong et al [12] use one host located in Texas A&M University to probe 53 U.S. Educational web sites and find out that the delay-distance correlation of U.S. Education Network is 0.829. Ziviani et al [13] study the delay-distance correlation in North America and Western Europe based on LibWeb and RIPE datasets. LibWeb dataset uses one probing host located in Paris to probe 135 university websites across the world. RIPE dataset uses 55 probing host located in U.S. and Europe to probe each other. The measurements show the delay-distance correlations in U.S. and Europe are 0.853 4 and 0.728 3，respectively, which indicates the delay-distance correlations in America and Europe network are all very strong. Dan Li measures delay between 240 probing hosts and 6000 websites in China and find that the delay-distance correlation of China Internet is only 0.373 4, much weaker than that of U.S. and Europe [11]. These works get the delay-distance correlation of several important networks in the world, which are very useful in many wide-area applications. However, it is hard for them to find rich-connected sub-networks because the classical definition of delay-distance correlation they used can only calculate the value. It cannot explain what factors and how these factors influence the value of delay-distance correlation.

To answer this question, this paper proposes a new delay-distance model (RTD-Corr model). It builds the relationship between delay-distance correlation and three main influencing factors: the tortuosity of the network path, the ratio of propagation delay and the direct geographical distance. Based on RTD-Corr model, we find that in general, as the differences between each $T$ (or the differences between $R$) become smaller, delay-distance correlation become stronger. Then based on the actual network characteristics, this paper discusses how to find rich-connected networks inside an actual poor-connected network, and points out that hosts inside one ISP and located in Regional Network Centers are more likely to be rich-connected sub-networks. Then based on the above analysis, a large-scale measurement is carried out to find rich-connected sub-networks. At last, based on founded rich-connected sub-networks, two classical IP geolocation algorithms (GeoGet & CBG) are modified [6,7].

## 1　RTD-Corr Model

### 1.1　Classical Definition

The classical definition of the delay-distance correlation is the first-order linear correlation coefficient between network delay and direct geographical distance. To calculate the delay-distance correlation of one network, researchers need to collect the dataset of (delay, distance) between probing hosts and landmarks in the network. Delay is the smallest RTT (round-trip delay) between a probing host and a landmark. Distance is the direct geographical distance between a probing host and a landmark. Assume the variance of delay is $V_{\text{delay}}$, the variance of distance is $V_{\text{distance}}$, and the covariance of delay and distance is cov(delay, distance), and Corr (delay-distance correlation) is calculated by the following formula [7]:

$$\text{Corr} = \frac{\text{cov}(\text{delay},\text{distance})}{\sqrt{V_{\text{delay}} \cdot V_{\text{distance}}}} \qquad (1)$$

The range of Corr is [−1,1]. In the area of IP geolocation, Corr is strong if it is beyond 0.7 and weak if under 0.7 because classical IP geolocation algorithms can perform well in a network whose Corr is beyond 0.7. Generally delay becomes larger as distance increases so that in most of the time Corr is positive. A negative Corr is treated as a very weak Corr because the accuracy of IP geolocation techniques is seriously affected in this kind of network.

### 1.2　Building RTD-Corr Model

The classical definition of delay-distance correlation can be used in measuring or calculating the value of delay-distance correlation of a specific network. However, it cannot explain why the delay-distance correlation of a network is strong or weak. In this section, we pro-pose a new model of delay-distance correlation based on the classical definition and the actual network factors.

Delay in the classical definition mainly consists of the propagation delay and the routing delay [13]. The propagation delay is the time that packets travel through the links between routers. The routing delay is the time a packet spends inside a router, mainly consisting of queuing time and processing time. There are other kinds of delays in delay (RTT) but they are too small compared with these two kinds [13], so we neglect them in our analysis.

Assume delay (or the whole delay) in the classical definition is $t_{whole}$, the propagation delay is $t_{propagation}$ and the routing delay is $t_{routing}$, then we can get the following formula:

$$t_{whole} = t_{propagation} + t_{routing} \quad (2)$$

Assume the ratio between the whole delay and the propagation delay is $R$, which is called the ratio of propagation delay, then we can get the following formula:

$$R = \frac{t_{whole}}{t_{propagation}} = \frac{t_{propagation} + t_{routing}}{t_{propagation}} \quad (3)$$

The range of $R$ is $(1, +\infty)$. If $R$ is approaching 1, the whole delay mostly consisting of the propagation delay and the routing delay can be almost neglected; If $R$ is approaching infinity, the whole delay mostly consists of the routing delay and the propagation delay can be almost neglected.

Assume there is a direct path between a probing host and a landmark, which is called the ideal link. The geographical distance of the ideal path is $D$ while the geographical distance of real ink between a probing host and a landmark is $d_{tortuosity}$, then the tortuosity of the network path $T$ is:

$$T = d_{tortuosity} / D \quad (4)$$

The range of $T$ is $[1, +\infty)$. If $T$ is approaching 1, the real link is near to the ideal link and is more direct; if $T$ is approaching infinity, the real link is more circuitous.

Assume the time that the packets traveling through the ideal link is $t_{ideal}$, and the traveling speed is $v$ ($v$ is a constant in this paper), then we can get the following formulas:

$$t_{ideal} = D / v \quad (5)$$

$$t_{propagation} = d_{tortuosity} / v \quad (6)$$

From formula (4), (5), (6), it is easy to know:

$$t_{propagation} = T \cdot t_{ideal} \quad (7)$$

From formula (5), (7), $t_{propagation}$ can be represented by $D$,

$$t_{propagation} = (T / v) \cdot D \quad (8)$$

Then based on (2),(3),(8), delay in the classical definition can be represented by $R$, $T$ and $D$:

$$\begin{aligned} t_{whole} &= t_{propagation} + t_{routing} \\ &= R \cdot t_{propagation} \\ &= \frac{R \cdot T \cdot D}{v} \end{aligned} \quad (9)$$

distance in the classical definition is $D$, so the Corr can be calculated by $R$, $T$ and $D$ as follows.

$$\begin{aligned} \text{Corr} &= \frac{\text{cov}(t_{whole}, D)}{\sqrt{V(t_{whole}) \times V(D)}} \\ &= \frac{\text{cov}(R \cdot T \cdot D, D)}{\sqrt{V(R \cdot T \cdot D) \cdot V(D)}} \end{aligned} \quad (10)$$

Assume $R$, $T$ and $D$ are independent from each other, the numerator can be rewritten as follows:

$$\begin{aligned} &\text{cov}[R \cdot T \cdot D, D] \\ &= E(R \cdot T \cdot D \cdot D) - E(R \cdot T \cdot D) \cdot E(D) \\ &= E(R \cdot T) \cdot E(D^2) - E(R \cdot T) \cdot E^2(D) \\ &= E(R \cdot T) \cdot V(D) \end{aligned} \quad (11)$$

Denominator can be rewritten as follows:

$$\begin{aligned} &V(R \cdot T \cdot D, D) \\ &= E[(R \cdot T \cdot D)^2] - E^2(R \cdot T \cdot D) \\ &= E[(R \cdot T)^2 - E(D^2)] - E^2(R \cdot T) \cdot E^2(D) \\ &= [V(R \cdot T) + E^2(R \cdot T)] \cdot E(D^2) - E^2(R \cdot T) \cdot E^2(D) \\ &= V(R \cdot T) \cdot E(D^2) + E^2(R \cdot T) \cdot V(D) \end{aligned} \quad (12)$$

From formula (10), (11), (12), we get the RTD-Corr model as (13) or (14).

$$\begin{aligned} \text{Corr} &= \frac{E(R \cdot T) \cdot V(D)}{\sqrt{V(R \cdot T) \cdot E(D^2) + E^2(R \cdot T) \cdot V(D)} \sqrt{V(D)}} \\ &= \frac{1}{\sqrt{\dfrac{V(R \cdot T) \cdot E(D^2) + E^2(R \cdot T) \cdot V(D)}{E^2(R \cdot T) \cdot V(D)}}} \\ &= \frac{1}{\sqrt{\dfrac{\{E[(R \cdot T)^2] - E^2(R \cdot T)\} \cdot E(D^2) + E^2(R \cdot T) \cdot [E(D^2) - E^2(D)]}{E^2(R \cdot T) \cdot [E(D^2) - E^2(D)]}}} \\ &= \frac{1}{\sqrt{\dfrac{E[(R \cdot T)^2] \cdot E(D^2) - E^2(R \cdot T) \cdot E^2(D)}{E^2(R \cdot T) \cdot E(D^2) - E^2(R \cdot T) \cdot E^2(D)}}} \end{aligned}$$

(13)

Formula (13) can also be shown as follows.

$$\mathrm{Corr} = \sqrt{\frac{E^2(R \cdot T) \cdot E(D^2) - E^2(R \cdot T) \cdot E^2(D)}{E[(R \cdot T)^2] \cdot E(D^2) - E^2(R \cdot T) \cdot E^2(D)}}$$

(14)

Formula (14) shows the relationship between delay-distance correlation and the actual network factors: the tortuosity of the network path, the ratio of propagation delay and the direct geographical distance.

### 1.3 Analyzing RTD-Corr Model

From RTD-Corr model, we can see the tortuosity of the network path, the ratio of propagation delay and the direct geographical distance of one specific network determines the value of the delay-distance correlation. This section will analyze how the delay-distance correlation changes as the three factors change.

From formula (14), we can see there are only slight differences between numerator and denominator of RTD-Corr model: the minuend of numerator is $E^2(R \cdot T) \cdot E(D^2)$ while the minuend of denominator is $E[(R \cdot T)^2] \cdot E(D^2)$. We already know the range of Corr is $[0,1]$, $R$ is $(1,+\infty)$, $T$ is $[1,+\infty)$ and $D$ is $(0,+\infty)$, so we can get the following relationship:

1) $E^2(R \cdot T) = E[(R \cdot T)^2]$, $\mathrm{Corr} = 1$;
2) $E^2(R \cdot T) \ne E[(R \cdot T)^2]$, $0 \wedge \mathrm{Corr} \wedge 1$;
3) $E^2(R \cdot T) \square E[(R \cdot T)^2]$, $\mathrm{Corr} \approx 0$;
4) $E(D^2) = E^2(D)$, $\mathrm{Corr} = 0$.

So we can conclude that the relationship of delay-distance correlation and actual network factors:

1) If all the $T$ (the tortuosity of the network path) are the same and all the $R$ (the ratio of propagation delay) are the same, delay-distance correlation of a network reaches the strongest (1 means the strongest);

2) If $E^2(R \cdot T) \square E[(R \cdot T)^2]$ (each $T$ is very different from others or each $R$ is very different from others), delay-distance correlation of a network reaches the weakest (0 means the weakest);

3) In general, as the differences between each $T$ (or the differences between each $R$) become smaller (larger), delay-distance correlation become stronger (weaker).

Here we do not consider the condition when $D$ is all the same, because it has little value in IP geolocation.

## 2 Rich-Connected Sub-Networks

Section 1 shows how the value of delay-distance correlation is affected by actual network factors. This section will discuss about how to find a sub-network whose delay-distance correlation may be stronger than the whole network. In this section, we take China Internet as an example because previous work has already validated that its network is poor-connected and classical IP geolocation algorithms cannot work so well in this network [11].

### 2.1 Derived Definitions of Delay-distance Correlation

Previous work only calculate one delay-distance correlation of a whole network (usually a country or a continent). To find rich-connected sub-networks, we need to calculate delay-distance correlation of different groups of hosts, which are inside one network. Thus, we need to introduce some derived definitions of delay-distance correlation which are used in this paper.

**Intra-ISP Corr & Inter-ISP Corr**. If the probing hosts and the landmarks are all in the same ISP, the Corr is an Intra-ISP Corr. If all the probing hosts belong to one ISP while the landmarks are all in another ISP, the Corr is an Inter-ISP Corr.

**Overall Corr of a certain network & the Corr of a probing host.** If the dataset of (delay, distance) includes all probing hosts' data of a network, the Corr is an overall Corr of this network. If the dataset of (delay, distance) comes from only one probing host, the Corr is referred as this probing host's Corr or the Corr of this probing host. However, if we do not point out the specific name of the probing host, Corr usually means the Overall Corr of a certain network.

For example, the Corr between China Unicom probing hosts and CERNET landmarks is Unicom-CERNET-Inter-ISP Corr, which is also the overall Unicom-CERNET -Inter-ISP Corr; the Corr between a specific China Telecom probing host and all China Telecom landmarks is the Telecom-Intra-ISP Corr of this probing host.

### 2.2 Possible Rich-Connected Sub-Networks in China Internet

Rich-connected sub-networks can be found by analyzing which sub-networks may be much stronger than others. Based on the RTD-Corr model and the characteristics of China Internet, we think Intra-ISP Corr and the Corr of probing hosts in Regional Network Centers are more likely to be rich-connected sub-networks.

**Intra-ISP Corr.** In China Internet, the number of IXP (Internet eXchange Point) is much less than those in rich-connected networks such as U.S. [14]. In addition, they mainly locate in several cities such as Beijing, Shanghai and Guangzhou [14]. The links between two ISPs are generally more circuitous than those inside one

ISP. Especially for hosts inside one city but belonging to different ISPs, their link may have to go a long detour to the IXP located in Beijing or Shanghai. Therefore, the differences between $T$ of links across ISPs are generally much larger than those inside one ISP. Besides, limited by the capacity of IXP, the routing delay of some links between different ISPs may be much larger than others. So the differences between $R$ of links across ISPs are also generally much larger than those inside one ISP. Based on the two reasons and RTD-Corr model, we think the Intra-ISP Corr of China may be much stronger than Inter-ISP Corr. So rich-connected sub-networks inside one ISP should be found at first.

**The Corr of probing hosts in Regional Network Centers**. If the intra-ISP Corr is not strong enough, we may consider the Corr of some probing hosts in particular cities, especially those located in Regional Network Centers. Tian *et al* [15] find out that major Chinese ISPs are highly hierarchical following China's political districts. Some cities act like the hubs of certain political districts, which are in charge of transmitting the main intra-district network traffic, and most inter-district network traffic. These cities, referred as Regional Network Centers, are usually important economic and political centers of one region or even the whole country. Probing hosts, which are in these Regional Network Centers tend to have straighter data transmission paths to landmarks while the data transmission paths from probing hosts in other cities sometimes may have to take a detour to the Regional Network Centers. So the differences between $T$ of links from probing host in Regional Network Centers are generally smaller than those in other cities. Based on this reason and the RTD-Corr model, we think the Intra-ISP Corr of probing hosts in Regional Network Centers may be much stronger than others. So the delay-distance correlation of the probing hosts in Regional Network Centers should also be measured.

### 2.3 Measurement of China Internet

In this part, we carry out network measurements to find rich-connected sub-networks in China Internet. The dataset of China Internet consists of round-trip time (RTT) and direct geographic distance between 90 probing hosts and 450 landmarks. RTT of each (probing host, landmark) pair is measured every minute for one week and only the minimum one is selected. The distance of each (probing host, landmark) pair is calculated by Vincenty's formula based on [16]. Mainland China consists of 31 provinces (or provincial administrative regions). The major ISPs which can cover the whole country are China Telecom, China Unicom, China Mobile and China Education and Research Network (CERNET). We manually deploy 90 probing hosts: 24 probing hosts belong to CERNET, 36 probing hosts belong to China Telecom and 30 probing hosts belong to China Unicom. The probing hosts cover 43 cities and 28 provinces of China, and most of them are in or near to Regional Network Centers. There are more than one probing hosts in certain cities. 450 landmarks are distributed evenly in three major ISPs and 30 province capitals, which means there are 5 landmarks in each provincial capital and each ISP. All landmarks are Web-based landmarks [4].

The overall Corr of whole China Internet (between 90 probing hosts and 450 landmarks) is 0.167 4, which is very weak and shows China Internet is a typical poor-connected network as a whole.

The Intra-ISP Corr and Inter-ISP Corr of three major ISPs in China are shown in Table 1. The first column of Table 1 represents the ISP information of the probing hosts and the first row represents the ISP information of the landmarks. Each number along the diagonal of Table 1 is the Intra-ISP Corr of one ISP. The other numbers are the Inter-ISP Corr between two ISPs. For example, the Intra-ISP Corr of China Telecom (Telecom-Intra-ISP Corr) is 0.6701 and the Inter-ISP Corr between China Telecom and CERNET (Telecom-CERNET-Inter-ISP Corr) is 0. 1398.

From Table 1, we can see all three Intra-ISP Corrs are between 0.6 and 0.7, which is close to rich-connected networks and much stronger than the overall Corr of China Internet.

**Table 1   The Intra-ISP Corr and Inter-ISP Corr of three major ISPs in China**

| ISP | CERNET | Telecom | Unicom |
|---|---|---|---|
| CERNET | 0. 689 5 | 0. 139 8 | 0. 029 8 |
| Telecom | 0. 161 4 | 0. 670 1 | 0. 314 2 |
| Unicom | 0. 029 6 | 0. 260 1 | 0. 611 4 |

Table 2 shows the Intra-ISP Corr and Inter-ISP Corr of CERNET probing hosts in different cities. The first column means the city where the probing hosts locate; The second column means the CERNET-Intra-ISP Corr of probing hosts; the third column means the CERNET-Telecom-Inter-ISP Corr of probing hosts; The fourth column means the CERNET-Unicom-Inter-ISP Corr of probing hosts.

From Table 2, we can see the CERNET-Intra-ISP Corr of probing hosts in 7 out of 8 cities are larger than 0.7. For inter-ISP Corr, the CERNET-Telecom-inter-ISP

Corr of probing hosts in 2 cities are larger than 0.7. And the Corrs of the probing hosts which are in Regional Network Centers (Beijing and Shanghai) or near to Regional Network Centers (Hangzhou) are usually stronger than those far away from Regional Network Centers (Chongqing).

**Table 2　The Intra-ISP Corr and Inter-ISP Corr of probing hosts in different cities of China (CERNET)**

| City | Intra-ISP | Inter-ISP(T) | Inter-ISP(U) |
| --- | --- | --- | --- |
| Beijing | 0.906 4 | 0.474 5 | 0.783 4 |
| Shanghai | 0.950 3 | 0.490 6 | 0.154 9 |
| Chongqing | 0.452 4 | −0.296 4 | −0.291 0 |
| Hangzhou | 0.926 7 | 0.599 6 | 0.714 0 |
| Wuhan | 0.767 1 | 0.367 0 | 0.176 0 |
| Chengdu | 0.782 1 | −0.437 5 | −0.131 3 |
| Jinan | 0.861 1 | 0.018 2 | 0.514 3 |

The data in Table 2 are only from probing hosts which belong to CERNET. We also calculate the Intra-ISP Corr and Inter-ISP Corr of probing hosts in the other two major ISPs. For all probing hosts in these three major ISPs of China, the Intra-ISP Corr of 51% of probing hosts are beyond 0.7 and the Inter-ISP Corr of 12% of probing hosts are beyond 0.7. Overall, the Corr of about 25% of the sub-networks in China Internet is beyond 0.7.

# 3　Modification of IP Geolocation Algorithms Based on Rich-Connected Sub-Networks

In this section, we modify two classical delay-based IP geolocation algorithms (GeoGet and CBG) based on the rich-connected sub-networks founded in Section 2 to check if their accuracy can be increased [10,11].

## 3.1　Modified GeoGet

3.1.1 How to Modify GeoGet

GeoGet maps a target host to the landmark which has the shortest delay to the target host. Its accuracy is based on the "shortest-closest" rule: the shortest delay comes from the closest distance [11]. The "shortest-closest" rule is closely related to delay-distance correlation. If the delay-distance correlation is strong, delay will increase as distance increases, so the probability that shortest delay comes from a landmark in the same city is high. If the delay-distance correlation is weak, distance has little impact on delay, the shortest delay may come from a landmark far away in other cities.

GeoGet needs the target host to probe landmarks to get the shortest delay. So its accuracy is actually determined by the delay-distance correlation between the target host and all landmarks. In China Internet, the intra-ISP Corr of 51% of probing hosts are beyond 0.7 and much stronger than the overall Corr of China Internet (0.1674). So we modify GeoGet by deploying and selecting landmarks in the same major ISP as target hosts.

The process of Modified GeoGet is:

1) deploy landmarks which can cover all major ISPs in each city;

2) cities are divided into different areas and there is one center city in each area;

3) check the ISP information of the target host and send landmarks which are in the same ISP in all center cities to the target host to probe;

4) choose the areas whose landmarks have the shortest delay as the candidate areas and send landmarks which are in the same ISP in all cities of the candidate areas to the target host to probe;

5) choose the city whose landmarks have the shortest delay as the city-level location of the target host.

3.1.2　Experiments

To examine the performance of Modified GeoGet, Mainland China is divided into 30 areas. 450 landmarks are deployed in 30 provincial capital cities, covering three main ISPs of China (these landmarks are actually the set of landmarks used in Section 2). Three landmarks are assigned in each non-capital city (in most non-cities there is one landmark in each major ISP, but we have to give up some cities for we cannot find enough reliable landmarks in each ISP). We manage to find 116 probing hosts in China Internet to play the role of target hosts.

To compare with the performances of Modified GeoGet, Original GeoGet make target hosts select landmarks which belong to the other two ISPs in step 3 and step 4. The rest steps are similar to Modified GeoGet.

We calculate the ratio of target hosts which are mapped to correct cities. When selecting 1 candidate areas, 5 landmarks (in the same ISP as target hosts) in each provincial capital city and 1 landmark (in the same ISP as target hosts) in each non-capital city, Modified GeoGet can successfully map 112, about 97% of target hosts to correct cities. Original GeoGet can map 44, about 38% of target hosts to correct cities when selecting 1 candidate area, 5 landmarks (in the other two ISPs) in each provincial capital city and 1 landmark (in the other two ISPs) in each non-capital city. So the city-level map-ping accuracy of GeoGet can be increased nearly 2 times based on the rich-connected sub-networks.

## 3.2 Modified CBG

### 3.2.1 How to Modify CBG

CBG measures delay and distance between each probing host and calculates each probing host's relationship of delay-distance conversion by drawing a tightest lower linear bound of all the (delay, distance) pairs, which is called bestline [10]. Then probing hosts measure delay to the target host and convert delay to distance using their bestlines. Every probing host draws a circle with center at their respective location and of radius equal to their own estimated distances. CBG believes that in most cases these estimated distances are larger than the actual distances so if all circles can intersect to a region, the location of the target host lies in the centroid of the region.

CBG is actually based on the assumption that Corr between a probing host and landmarks is strong. If the Corr between a probing host and target hosts is strong, the bestline can get an accurate estimated distance. If the Corr is weak, the distance tends to be overestimated. For example, during the measurement of China Internet in last section, we find that the CERNET-Intra-ISP Corr of probing host **pl2.6test.edu.cn** is 0.905 6 while the CERNET-Telecom-inter-ISP Corr of probing host **pl2.6test.edu.cn** is only −0.038 6 (in city-level IP Geolocation, a minus Corr can be seen as a very weak Corr like 0). When using **pl2.6test. edu.cn** to estimate distance to landmarks belonging to Telecom, the error distance is much larger than those belonging to CERNET as shown in Fig.1.

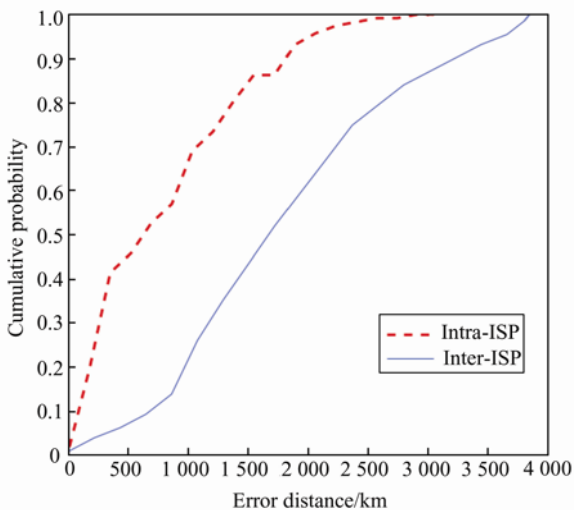

**Fig.1 CDF of the difference between estimated distance and actual distance using bestline of intra-ISP and inter-ISP(delay,distance) pairs**

To get more accurate results, Modified CBG should select the probing hosts and landmarks which are in the same rich-connected sub-network with target hosts. In China Internet, the Intra-ISP Corr of 51% of probing hosts are beyond 0.7 and the Inter-ISP Corr of 12% of probing hosts are beyond 0.7. We can modify CBG based on a Corr-based probing host selection strategy: Modified CBG should choose probing hosts which are in the same ISP with target hosts and have an Intra-ISP Corr more than 0.7 or probing hosts which are in the other ISPs and have an Inter-ISP Corr more than 0.7.

The process of Modified CBG is as follows:

1) deploy probing hosts belonging to all major ISPs in each city and measure delay and distance between probing hosts;

2) calculate the intra-ISP Corr and inter-ISP Corr of each probing host;

3) plot Intra-ISP and Inter-ISP bestlines for each probing host;

4) check ISP information of the target host, choose probing hosts which are in the same ISP and have an intra-ISP Corr more than 0.7;

5) if there is no eligible probing host in one city, choose probing hosts which are in the other ISPs and have an inter ISP Corr more than 0.7;

6) probing hosts which are in the same ISP use intra-ISP bestlines to estimate distance while probing hosts in the other ISPs use inter-ISP bestlines to estimate distance.

7) the next steps are similar to Original CBG as introduced in the start of this section.

### 3.2.2 Experiments

We geolocate 450 target hosts which cover 30 provinces and 3 major ISPs of China (they are actually the 450 landmarks used in Section 2). The probing hosts used in experiments are the 90 probing hosts used in Section 2. To compare the mapping accuracy of Modified CBG with Original CBG, a contrast group of probing hosts is constructed for Original CBG. Modified CBG selects one probing host in each city based on the Corr-based selection technique while Original CBG uses a contrast group of probing hosts by randomly selecting one probing host in each city.

There are 28 target hosts that Modified CBG and Original CBG cannot form intersection regions. Figure 2

shows the CDF (cumulative distribution function) of error distances of Modified CBG and Original CBG. In Modified CBG, the median error distance is 315.4 km while Original CBG is 629.8 km. The average error distance in Modified CBG is 544.1 km while Original CBG is 712.6 km. The comparison results show that the Modified CBG can reduce the median error distance of Original CBG by about 50%.

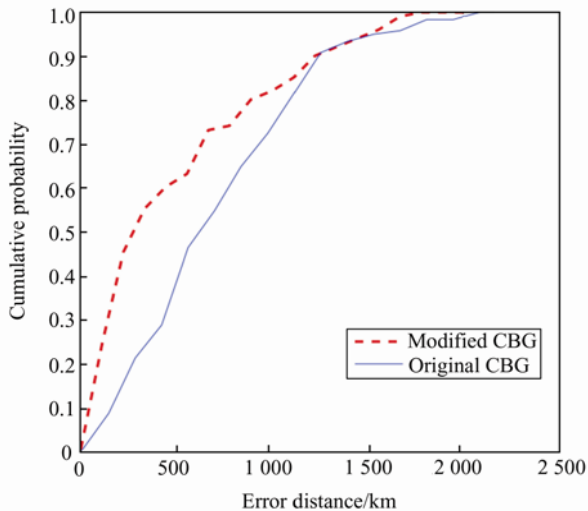

**Fig.2   CDF of the error distances of Modified CBG and Original CBG**

## 4   Conclusion

IP geolocation is important to location-based security services. Although classical delay-based IP geolocation algorithms work well in rich-connected networks, their accuracy still need to be improved in poor-connected networks. This paper first proposes a new delay-distance correlation model (RTD-Corr model). It shows that as the differences between each $T$ (or the differences between each $R$) become smaller (larger), delay-distance correlation becomes stronger (weaker). Based on the RTD-Corr model and the actual network characteristics, this paper points out that Intra-ISP Corr and the Corr of probing hosts in Regional Network Centers are more likely to be rich-connected sub-networks in China Internet. Then we carry out network measurements and find out about 25% of sub-networks in China Internet are rich-connected. Based on the founded sub-networks, we modify two classical delay-based IP geolocation algorithms, GeoGet and CBG. The experiments show the mapping accuracy of GeoGet can be increased by nearly two times and the median error distance of CBG can be reduced by about 50%. In further research, we will try to design more accurate IP geolocation algorithms suitable to poor-connected networks.

□